\documentclass[12pt]{iopart}
\usepackage{graphicx}
\graphicspath{{./figs/}}
\begin{document}

\title{SNEWS: The SuperNova Early Warning System}


\author{
Pietro Antonioli\ddag,
Richard Tresch Fienberg\S,
Fabrice Fleurot$\|$,
Yoshiyuki Fukuda\P,
Walter Fulgione$^+$,
Alec Habig*,
Jaret Heise$\sharp$,
Arthur B McDonald$^a$,
Corrinne Mills$^{b,f}$,
Toshio Namba$^c$,
Leif J Robinson\S,
Kate Scholberg$^d$ 
\footnote[1]{To
whom correspondence should be addressed \eads{schol@mit.edu}},
Michael Schwendener$\|$,
Roger W Sinnott\S,
Blake Stacey$^d$,
Yoichiro Suzuki$^c$,
R\'{e}da Tafirout$\|$$^g$,
Carlo Vigorito$^+$,
Brett Viren$^e$, 
Clarence Virtue$\|$, and
Antonino Zichichi\ddag
}

\address{\ddag\ University of Bologna and INFN Bologna, Via Irnerio 46, 40126 Bologna, Italy }
\address{\S\ \textit{Sky \& Telescope}, 49 Bay State Rd, Cambridge, MA 02138-1200, USA}
\address{$\|$\ Department of Physics and Astronomy, Laurentian University, Sudbury, Ontario P3E 2C6 Canada}
\address{\P\ Department of Physics, Miyagi University of Education, Sendai, Miyagi 980-0845, Japan}
\address{$^+$  IFSI-CNR Torino, University of Torino and INFN Torino, 
Via P.Giuria 1, 10125 Torino, Italy}
\address{*\ Department of Physics, University of Minnesota, Duluth, MN 55812-2496, USA}
\address{$\sharp$\ Los Alamos National Laboratory, Los Alamos, NM 87545}
\address{$^a$ Department of Physics, Queen's University, Kingston, Ontario K7L 3N6, Canada}
 \address{$^b$ Department of Physics, Boston University, Boston, MA 02215, USA}
\address{$^c$ Institute for Cosmic Ray Research, University of Tokyo, Kashiwa,
Chiba 277-8582, Japan}
\address{$^d$ Department of Physics, Massachusetts Institute of Technology, Cambridge, MA 02139, USA}
\address{$^e$ Brookhaven National Laboratory, P. O. Box 5000, Upton, NY 11973-5000}
\address{$^f$ Current address: Department of Physics, University of California, Santa Barbara, Santa Barbara, CA 93106 USA}
\address{$^g$ Current address: Department of Physics, University of Toronto,
Toronto, Ontario M5S 1A7, Canada}

\begin{abstract}
This paper provides a technical description of the SuperNova
Early Warning System (SNEWS), an international network of experiments
with the goal of providing an early warning of a galactic
supernova.
\end{abstract}

\maketitle

\section{Introduction}

The famous supernova SN1987A in the Large Magellanic Cloud (LMC) brought 
the field of supernova neutrino astrophysics to life.  Two water 
Cherenkov detectors, Kamiokande II and IMB, detected 20 events between 
them~\cite{Hirata1,Bionta,Bratton,Hirata2}; two scintillator detectors, Baksan 
and LSD~\cite{Alekseev,Aglietta} also reported observations. The sparse 
SN1987A neutrino data were sufficient to confirm the baseline model of 
gravitational collapse causing type II SNe and to put limits on neutrino 
properties (such as a $\bar{\nu}_e$ mass limit of around 20~eV.)  To 
make distinctions between different theoretical models of core collapse 
and supernova explosions and to extract more information about neutrino 
properties, we await the more copious neutrino signal which the new 
generation of large neutrino experiments will detect from the next such 
event in our Galaxy. 

When the core of a massive star at the end of its life collapses, less 
than 1\% of the gravitational binding energy of the neutron star 
will be released in the forms of optically visible radiation and 
the kinetic energy of the expanding remnant. The remainder of the binding 
energy is radiated in neutrinos, of which $\sim$1\% will be electron 
neutrinos from an initial ``neutronization'' burst and the remaining 
99\% will be neutrinos from the later cooling reactions, roughly 
equally distributed among flavors.  Average neutrino energies are 
expected to be about 13-14~MeV for $\nu_{e}$, 14-16~MeV for $\bar{\nu}_e$, 
and 20-21~MeV for all other flavors.  The neutrinos are emitted over a 
total timescale of tens of seconds, with about half emitted during the 
first 1-2~seconds.  Reference~\cite{Burrows} summarizes the expected 
features of a core collapse neutrino signal; more recent simulation 
work can be found in {\it e.g}.~\cite{Thompson, Mezzacappa}. 

A core-collapse supernova in our Galaxy will bring a wealth of
scientific information.  The neutrino signal will provide information
about the properties of neutrinos themselves and astrophysicists will
learn about the nature of the core collapse.  One
unique feature of the neutrino signal is that it is \textit{prompt} ---
neutrinos emerge on a timescale of tens of seconds, while the first
electromagnetic signal may be hours or days after the stellar collapse.  
Therefore, neutrino observation can provide an \textit{early alert}
that could allow astronomers a chance to make unprecedented
observations of the very early turn-on of the supernova light curve;
even observations of SNe as young as a few days are rare
for extra-galactic supernovae.  The environment immediately
surrounding the progenitor star is probed by the initial stages of the
supernova.  For example, any effects of a close binary companion upon
the blast would occur very soon after shock breakout.  UV and soft
x-ray flashes are predicted at very early times.  Finally, there may
be entirely unexpected effects---no supernova has ever been observed
very soon after its birth.  Although the neutrino signal will be
plentiful in practically all galactic core collapses, it is possible that there will be
little or no optical fireworks (the supernova ``fizzles''); the nature
of any observable remnant would then be very interesting.

This paper focuses on the prompt alert which is possible using
the neutrino signal.  We will describe the technical aspects of the 
system.  

Section~\ref{overview} gives an overview of SNEWS, and
Section~\ref{signal} briefly covers the expected signal in current
detectors.  Section~\ref{3ps} discusses some issues associated with
SNEWS.  Section~\ref{indiv} introduces the individual
experiments' monitors.  Section~\ref{implementation} covers SNEWS
implementation and defines the coincidence conditions and alert
scheme.  Section~\ref{highrate} describes the results of the ``high-rate''
system test performed in 2001.  Section~\ref{alert} describes the alert to
the astronomical community.  Section~\ref{future} gives future
directions. The final section summarizes.
 
\section{SNEWS Overview}\label{overview}

The SNEWS (SuperNova Early Warning System) collaboration is an
international group of experimenters from several 
supernova neutrino-sensitive experiments.  The primary goal of SNEWS
is to provide the astronomical community with a prompt alert for a
galactic supernova.  An additional goal is to optimize global
sensitivity to supernova neutrino physics, by such cooperative work as
downtime coordination.

The idea of a blind central coincidence computer receiving signals
from several experiments has been around for some time
({\it e.g.}~\cite{snwatch}.) In addition to the basic early warning
advantages of a neutrino detector, there are several benefits from a
system involving neutrino signals from two or more different
detectors.  First, if the supernova is distant and only weak signals
are recorded, a coincidence between signals from different detectors
effectively increases the sensitivity by allowing reductions in alarm
thresholds and allowing one to impose a minimum of (possibly
model-dependent) expectations on the form of the signal.  Second, even
if a highly sensitive detector such as Super-K is online,
\textit{requiring a coincidence among several detectors effectively
reduces the ``non-Poissonian'' background present for any given
detector and enormously increases the confidence in an
alert.}\footnote{``Non-Poissonian'' refers to background alarms whose
rate cannot be well predicted according to a constant-background-rate
Poisson distribution. Detector effects such as flashing phototubes and
electronics problems fall under this category.  Rates may also be
locally Poissonian, just non-stationary.}  Background alarms at widely
separated laboratories are highly unlikely to be correlated.  Without
the additional confidence from coincident neutrino observations, it
would be very difficult for any individual detector to provide an
\textit{automated} alert to astronomers.  Finally, using signals from
more than one detector, there is some possibility for determining the
direction of the source when a single detector alone can provide no
information (see reference~\cite{pointing}.)  Unfortunately triangulation is
in practice quite difficult to do promptly, and cannot point as well
as individual detectors.  

An important question for SNEWS is: how often is a galactic supernova
likely to occur?  Estimates vary widely, but are typically in the
range of about one per 30 years ({\it e.g.} \cite{snfreq}.)  This is
frequent enough to have a reasonable hope of observing one during the
next five or ten years, but rare enough to mean that we must take
special care not to miss anything when one occurs.

The charter member experiments of SNEWS are Super-Kamiokande (Super-K)
in Japan, the Sudbury Neutrino Observatory (SNO) in Canada and the
Large Volume Detector (LVD) in Italy\footnote{MACRO\cite{macro} was
another charter member, and was involved with SNEWS until it turned
off in 2000.}.  Representatives from AMANDA, IceCube, KamLAND,
Borexino, Mini-BooNE, Icarus, OMNIS, and LIGO participate in the SNEWS
Working Group, and we hope will eventually join the active
coincidence.

There is currently a single coincidence server, hosted by Brookhaven
National Laboratory.  We expect that additional machines will be
deployed in the future.  The BNL computer continuously runs a
coincidence server process, which waits for alarm datagrams from the
experiments' clients, and provides an alert if there is a coincidence
within a specified time window (10 seconds for normal running.)
We have implemented a scheme of ``GOLD'' and ``SILVER'' alerts:
GOLD alerts are intended for automated dissemination to the community;
SILVER alerts will be disseminated among the experimenters, and
require human checking.  

As of this writing, no inter-experiment
coincidence, real or accidental, has ever occurred
(except in high rate test mode), nor has any core collapse event been
detected within the lifetimes of the currently active experiments.
  
\section{The Supernova Signal and Current Detectors}\label{signal}

There are several classes of detectors capable of observing neutrinos
from gravitational collapse. Most supernova neutrino detectors are
designed primarily for other purposes, {\it e.g.} for proton
decay searches, solar and atmospheric neutrino physics, accelerator
neutrino oscillation studies, and high energy neutrino source
searches.

\begin{table}[t]
\begin{centering}
\begin{tabular}{||c|c|c|c|c|c||}\hline\hline
Detector type & Material & Energy & Time & Point & Flavor \\ \hline\hline
scintillator & C,H & y & y & n & $\bar{\nu}_e$\\ \hline
water Cherenkov &  H$_2$0 & y & y & y & $\bar{\nu}_e$ \\ \hline
heavy water & D$_2$0 & NC: n & y & n & all \\ \cline{3-6}
         &      & CC: y & y & y & $\nu_e$,$\bar{\nu}_e$\\ \hline
long string water Cherenkov& H$_2$O & n & y & n & $\bar{\nu}_e$ \\ \hline
liquid argon & Ar & y & y & y & $\nu_e$ \\ \hline
high Z/neutron  & Pb, Fe & y & y & n & all \\ \hline
radio-chemical & $^{37}$Cl, $^{127}$I, $^{71}$Ga  & n & n & n & $\nu_e$ \\
 \hline \hline 
\end{tabular}
\caption{Supernova neutrino detector types and their primary capabilities.\label{tab:detector_types}}
\end{centering}
\end{table}

Table~\ref{tab:detector_types} gives a brief overview of the supernova
neutrino detector types.  More detailed information about supernova
detection capabilities can be found in reference~\cite{nu2k}.  To
summarize briefly: scintillator and water Cherenkov detectors are
sensitive primarily to $\bar{\nu}_e$; those with neutral current
capabilities (heavy water, high Z/neutron, and also water
Cherenkov and scintillator to some extent) are sensitive to all
flavors.  Water Cherenkov and heavy water detectors have significant
pointing capabilities.  All except radiochemical can see neutrinos in
real-time. All have energy resolution except long string water
Cherenkov and radiochemical.

Table~\ref{tab:specific_detectors} lists specific supernova neutrino
detectors and their capabilities
\cite{LVDref,superk,SNO,Amanda,icarus,miniboone,borexino,omnis,uno}.
For a summary of supernova neutrino capabilities of future detectors,
please see~\cite{whitepaper, ness}.
\footnote{Note that the currently running 
Super-K II (after reconstruction in 2002) has nearly
the same supernova sensitivity as Super-K I; a slight increase 
in energy threshold due to loss of phototubes will cause only a few percent
loss of total signal events.}
\footnote{Gravitational wave
detectors deserve some note here. Large interferometer experiments
such as LIGO, Virgo, GEO, TAMA and ACIGA\cite{gravwave} as well as
cryogenic antennas belonging to the IGEC
collaboration\cite{igec} may have the capability of detecting
gravitational wave signals from asymmetric supernova explosions
(although the details of a stellar collapse gravitational wave signal
are not yet well understood.)  When these detectors reach maturity
over the next several years, they will become an important part of a
stellar collapse network, and combined neutrino and gravitational wave
data will be an extremely valuable source of information for testing
supernova models.  The gravitational wave signal may be more prompt
even than the neutrino signal, and in fact, may provide a $t=0$ for a
neutrino time of flight mass measurement (see {\it
e.g.}~\cite{Arnaud}.)  The scientific potential from combined
gravitational wave and neutrino signals from stellar collapse is
exciting and largely unexplored territory.}

\begin{table}[htbp]
\begin{centering}
\begin{tabular}{||c|c|c|c|c|c||}\hline\hline
Detector & Type & Mass & Location  & \# of events & Status\\ 
         &      & (kton) &         & @8.5 kpc &  \\ \hline\hline
Super-K & H$_2$O Ch. & 32    & Japan  & 7000 & running \\ \hline
SNO & H$_2$O,& 1.4 & Canada &  300 & running \\
    & D$_2$O & 1   &        &  450 &  \\ \hline
LVD & scint. & 1 & Italy & 200 & running\\ \hline
AMANDA & long string  & M${\rm eff}\sim$0.4/pmt & Antarctica & & running \\ \hline 
Baksan & scint. & 0.33  & Russia  & 50 &  running\\ \hline
Mini-BooNE & scint. & 0.7  & USA  & 200  &  running\\ \hline
KamLAND & scint. & 1  &  Japan  & 300 &  running\\ \hline
Borexino & scint. & 0.3  & Italy  & 100 &  200x\\ \hline
Icarus & liquid argon & 2.4 & Italy & 200 & 200x \\ \hline
OMNIS  & high Z (Pb) & 2 &  USA & 2000 &  proposed \\ \hline
LANNDD & liquid argon & 70 & USA & 6000 & proposed\\ \hline 
UNO/Hyper-K  & H$_2$O Ch.  &  600-1000 & USA/Japan  & $>$100,000 &  proposed \\ \hline\hline
\end{tabular}
\caption{Specific supernova neutrino detectors.
The expected numbers of events are approximate, and
refer to yields in the dominant channels.\label{tab:specific_detectors}
}
\end{centering}
\end{table}

\section{The Three ``P'''s}\label{3ps}

In order to make the best use of a neutrino burst supernova alert, the
astronomical community needs the ``three P's'':  ``prompt'', ``pointing''
and ``positive''.  We comment on each of these below.

\subsection{``Prompt''} 

The alert must be as prompt as possible to
catch the early stages of shock breakout, which occurs within
hours (or less) of core collapse.  We estimate an alert dissemination time
of five minutes or less for an automated (GOLD) alert.  
A SILVER alert involving human-checked alarms would take longer,
optimistically 20~minutes or so, but
perhaps longer.

\subsection{``Pointing''} 

Clearly, the more accurately we can point to a core collapse event
using neutrino information, the more likely it will be that early
light turn-on will be observed by astronomers.  Even for the case when
no directional information is available ({\it e.g.} for a single
scintillator detector online) it is still useful for astronomers to
know that a gravitational collapse event has occurred. However any
pointing information at all is extremely valuable.  The question of
pointing to the supernova using the neutrino data has been examined in
detail in reference~\cite{pointing}.  There are two ways of pointing
with neutrinos: first, individual detectors can make use of asymmetric
reactions for which the products ``remember'' the direction of the
incoming neutrino.  Second, the timing of the neutrino signals in
several detectors can be used to do triangulation.
Reference~\cite{pointing} estimates roughly $5^\circ$ pointing accuracy
for Super-K and $20^\circ$ pointing accuracy for SNO, given a galactic
center core collapse.  Triangulation is less promising, and presents
practical difficulties: it requires immediate and complete exchange of
event-by-event information, which is difficult in practice, and we
do not plan to attempt it promptly.

We do not anticipate that SNEWS will disseminate pointing information
as part of the initial alert message in the short term
(although this may change); this information will come
from the individual experiments, and may not be available immediately.
Each experiment establishes its own protocol for making 
estimated pointing information available.

\subsection{``Positive'':}

There must be no false supernova alerts to the astronomical community.
A single experiment cannot realistically decrease the false alert rate
to zero, since there will always be some residual rate of false alerts
from Poissonian and non-Poissonian sources.  However, by requiring an
inter-experiment coincidence, the false alert rate can be decreased to
nearly zero: this is the great strength of SNEWS.  We have chosen the
nominal acceptable average false alert rate to be \textbf{one per century}.
The following section is devoted to the question of ensuring a
false alert rate which is sufficiently low.

\subsubsection{False Alerts}

The fundamental motivation for the SNEWS coincidence is the reduction of
false alerts.  We categorize the possibilities for false alerts below:

\begin{enumerate}

\item \textit{Accidental Coincidences}

Accidental (random) coincidences imply that there was no actual association
with an astronomical event and that the coincidence occurred by chance.
The rough expected rate of accidental coincidences can be calculated by
assuming equal, constant, uncorrelated alarm rates for each
experiment.  

Figure~\ref{fig:acccoinc1} shows the 
average interval between accidental alerts for an $n$-fold coincidence
of $N$ experiments, for a 10~second coincidence window and
an individual experiment background alarm rate of one per week.
This plot shows that an individual experiment alarm rate of one per week
is acceptable only if four or fewer experiments are online, or
if a three-fold coincidence if required; otherwise a lower individual
experiment rate is required.    Based on these considerations,
the requirement for an experiment to participate in SNEWS is an average 
alarm rate of \textbf{no more than 1 per week}.  We may adjust
the criteria defined in this paper if 
more than four experiments are running.

\begin{figure}[t]
\hspace{2cm}
\begin{center}
\includegraphics[width=10 cm]{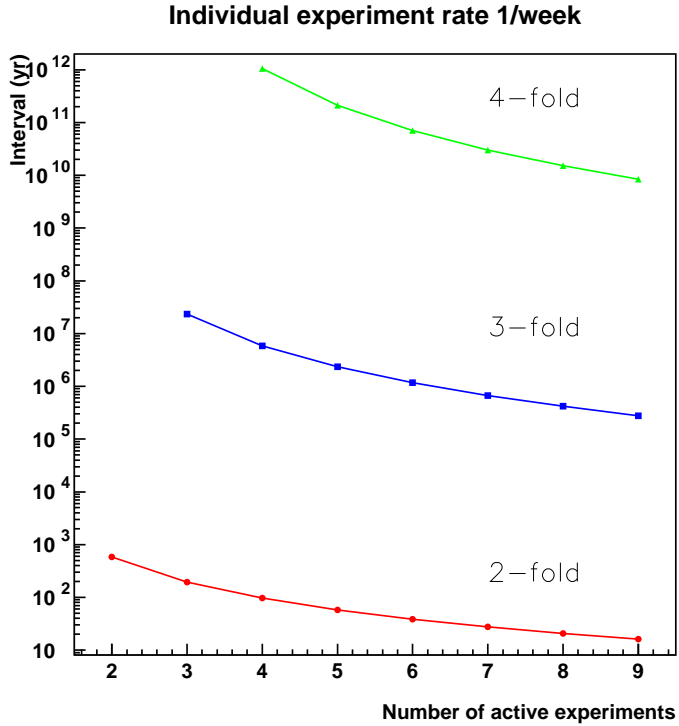}
\caption{\label{fig:acccoinc1}
Average interval between accidental alerts for an $n$-fold coincidence
of $N$ experiments, for a 10~second coincidence window and
a uniform individual experiment background alarm rate of one per week.}
\end{center}
\end{figure}

In reality, individual alarm rates are not strictly Poissonian and
change with time.  However, they may be Poissonian on shorter time
scales, and so long as instantaneous individual rates do not exceed a
certain value, the accidental coincidence rate can be made as small as
desired.  In Section~\ref{coinc} we detail how we deal with potentially
changing alarm rates.

\item \textit{Non-astrophysical Correlated Bursts}

The possibility of some correlation between bursts seen in the
individual detectors, which is not astrophysical in origin, exists to
the extent that some credible coupling can be shown to exist between
detectors. For participating detectors that may be physically close to
one another there are a large number of possible couplings from a
shared local environment (electrical noise, ambient pressure, seismic,
etc..) For participating detectors that are very well separated from
one another one has to invoke more fanciful and substantially less
probable couplings such as solar activity, solar flares, or widescale
upper atmospheric electrical disturbances. The most credible coupling
for separated detectors may well be the seismic one, but even that is
not really plausible.

\item \textit{Malicious Actions}

A fake alert sent to the astronomical community by hackers breaking in
to our machines is a remote possibility, but one which we view
seriously.  Breaching several machines at the individual experiments
and creating false alarms at the client datagram level would have the
same end effect, but would require more knowledge of the detailed and
widely different operation of several detectors, so we feel this is
much less of a concern.  To prevent malicious actions, we take a
serious approach to the security of the server and the connections to
it. We require that the server be housed at a national laboratory with
designated personnel to take responsibility for the security and
maintenance of the machine.

\end{enumerate}

\subsection{Privacy}

Another ``P'' (relevant to experimenters more than to astronomers) is
``Privacy''.  To satisfy inter-experiment privacy needs (and in
addition to help ensure secure data transmission), we have set up a
formal set of rules for data sharing and have structured the
collaboration around these rules.  The ``SNEWS subgroup'' is a working
group of a few people per experiment, designated by our Advisory Board
(spokespeople of the active experiments.)  Only subgroup members have
access to the alarm data from all experiments. Subgroup members agree
not to propagate information without explicit approval from the
Advisory Board.

\section{Description of Individual Experiments' Supernova Triggers}\label{indiv}

In this section we will describe briefly the online supernova
monitoring systems of Super-K, SNO and LVD, which provide alarm input
to the SNEWS coincidence.  The supernova capabilities of the detectors
are well known and details can be found elsewhere~\cite{nu2k}; other
details of the triggers, monitoring systems and analyses are also
described elsewhere~\cite{sk,sno,lvd}.  Although each of these
three experiments takes a somewhat different approach to real-time
monitoring, every one is sensitive to a galactic supernova and can
provide an alarm on a timescale of minutes.

A note on SNEWS terminology: an ``alarm'' refers
to a supernova neutrino burst candidate detected by an individual experiment,
according to conditions defined for that experiment.  An ``alert''
refers to a coincidence between alarms, and the conditions which
define an alert are described in section~\ref{implementation}.
This section describes the individual experiment
alarm conditions only.  The detailed neutrino event information comprising
the alarm bursts is not sent to the coincidence server.
 
\subsection{Super-K Supernova Online Monitor}


The Super-K supernova alarm system involves software that does a
prompt pre-analysis before full reconstruction.  Roughly two-minute
chunks of data called ``subruns'' are sent from the event builder via
``express-line'' to a dedicated supernova burst monitor machine,
skipping the usual steps required before data are sent
to the offline processes.  The low energy trigger
events are then searched for clusters in several time windows
(0.5 seconds, 2 seconds, 10 seconds.)

If the ``pre-multiplicity'' thresholds are exceeded for any of these
time windows, then the ``pre-candidate'' is passed to a second monitor
program for further analysis.  Standard noise reduction algorithms
similar to those applied for solar neutrino analysis~\cite{sk_solar}
are applied, and the search is performed again, this time on events
passing the cuts and with a higher energy threshold.  Full energy
reconstruction is not done (to save time), but vertices and directions
are reconstructed.  If any cluster passes a second pass multiplicity
threshold, the multiplicity $N$ in a 20~second time window is counted;
in addition, the mean distance between event positions in the cluster,
$R_{\rm mean}=\frac{1}{C}\sum_{i=1}^{N-1}\sum_{j=i+1}^{N} |
\vec{r_i}-\vec{r_j} |$, where $C$ is the number of pairs, is
calculated.  Background supernova alarms arise from muon spallation
events and ``flashing'' tubes.  Both of these types of fake clusters
have event vertex distributions which are highly non-uniform, and
which will yield small $R_{\rm mean}$ values.  Candidate clusters with
sufficiently high $N$ and $R$ are considered to be supernova
candidates.

Because reconstruction of thousands of events in a real supernova
burst ($\sim$ 5000~events) could require an hour or more to fully
analyze, pre-alarms are generated after 100~events if a candidate is
found (roughly a 1~minute timescale.)

If an alarm burst candidate is found, a datagram is sent to the SNEWS
server, and shiftworkers are alerted. Detailed information about the
candidate is made available to shiftworkers (present 24 hours a day
onsite.)  The shiftworker checks for the existence of spallation
muons, examines reconstructed vertices and their goodness, and also
checks the exploded view of the PMT hit pattern.  A preliminary
estimate of the supernova direction from elastic scattering is
available at this point.  If a good supernova candidate is identified,
an offline process will re-analyze to provide full, precise
reconstruction within a few hours.

\subsection{SNO Supernova Online Monitor}

At SNO, custom readout electronics collect
the PMT data underground and pass that information to an event builder.
Built events are then sent
to an event dispatcher process running on a surface
computer, which is used for online monitoring.
A fully detailed description will be found
elsewhere~\cite{sno}. Summarized here is the basic machinery
of the trigger, which consists of three distinct levels which are fast
and completely automated:

\begin{itemize}
\item {\it Level 1}: This is the burst monitor which 
looks in the datastream for a certain number of events above a
certain energy threshold within a certain time window. 
At present the multiplicity threshold is set to 30 events above
approximately 4 MeV in a 2 second time window which provides a
good sensitivity to a galactic supernova.
Dynamic thresholds
are used whenever calibration sources are introduced in the detector. 
Bursts satisfying the multiplicity criteria are written to a data file and then
transferred to an analysis machine.
\item {\it Level 2}: At this level events are calibrated and analyzed on an
event-by-event basis. The main task of this second-level trigger 
is to identify events with anomalous time and charge as well as
events with geometric signatures of particular detector pathologies
in order to cut them from the burst data set. For example, events with low
charge to number of hit ratio usually indicate electrical pickup. 
A set of data cleaning cuts are applied which are meant to reject
known instrumental background with a very high efficiency. Cherenkov
events pass those cuts with very little sacrifice.
\item {\it Level 3}: If more than 35\% of the events composing a burst
survive the data cleaning cuts, an alert is sent to the SNEWS server and
a dialout computer contacts the members of the SNO supernova trigger group.
In the meantime a more in-depth analysis is performed to extract 
fitted vertices and direction cosines. 
The relative event fractions occurring in the 
$\mathrm {D_2O}$/$\mathrm {H_2O}$ volumes are extracted and a search
algorithm uses the events' direction cosines to find the electron
scattering (ES) events, which are expected to best convey information
about the direction of the possible supernova. 
\end{itemize}

 The Level 2 analysis produces a set of histograms which are mainly 
useful for quick burst diagnostic by the operator and any 
interested party.  Besides hit and time distributions,
crate/slot/channel occupancies are provided, which are expected to be flat 
for a supernova signal.
The Level 3 analysis produces a set of histograms using fitted
vertices in both the light and heavy water volumes as well as angular
distribution of the events' fitted directions. Each burst is catalogued
and automatically archived on the SNO private WWW server.

\subsection{LVD Supernova Online Monitor}

In the LVD experiment the scintillator counting rate is continuously
monitored by a DAQ task, which examines all data collected in real
time.  A simple and fast muon rejection algorithm makes a
pre-selection of $\nu$-candidate signals, registered by the experiment
with a $12.5$~ns time precision.  This first selection level
does not apply cuts on pulse energy and topological distribution.\\ A
separate on-line monitor task looks for burst candidates from the
reduced data stream. The search algorithm is based on a pure
statistical analysis of the time sequence of events including some
additional cuts.  The code processes the sequence in order to
extract significant clusters of pulses having an expected frequency,
induced by the accidental background, lower than a predefined
threshold.\\ At this level, pulse energy is required to be in the
7-100 MeV range in order to avoid fluctuation effects at the edge of
energy threshold and problems due to electronic noise, as well as to
reject single counter muon signals.  After these cuts the background
pulse frequency is found to be very stable and corresponds, for the
full LVD configuration, to $f_B=0.2$~Hz.  The resulting $\nu$-pulse
candidate time sequence, collected inside a 1000-pulse deep circular
buffer, is processed by the alarm module of the monitoring
code. Buffered events are processed in fixed time windows $\Delta
T=20$~s originating at the start run time. For each asynchronous
window the number $N_\nu$ of contained single pulses is obtained.
Then the Poissonian probability $P_C$ to have $k\geq N_\nu$ events in
the cluster is calculated according to: $P_C(k \geq N_\nu,\lambda)=
\sum_{k=N_\nu}^\infty \frac{\lambda^k\cdot e^ {-\lambda}}{k!}$, where
$\lambda=f_B*20$ is the mean expected number of pulses due to
background rate.  To optimize sensitivity, the online frequency $f_B$
is evaluated each time a new pulse is inserted into the buffer. The
alarm threshold probability is obtained from the above expression by
fixing a global alarm frequency $f_A$.  The predictive capability of
the selection algorithm has been checked with real and simulated data
as a function of the required global alarm frequency ($f_A=1/{\rm
hour},...1/{\rm year}$.)\\ Finally, to reduce the number of false
alarms, for each selected candidate a topological check is
applied. For a real supernova burst candidate a uniform distribution
of pulses between involved counters is expected.  If not, counters
with abnormal high counting are excluded and the resulting cluster is
re-analyzed.  Surviving clusters are considered to be candidate
alarms, and corresponding datagrams are sent to the SNEWS servers; all
related information is saved for further analysis.  Online
event buffering and processing gives less than 2 minutes delay between
the burst time and the alarm notification.  The LVD shiftworker and
experts within the collaboration are notified.

\section{SNEWS Coincidence Implementation}\label{implementation}

This section describes the hardware setup and software developed for
SNEWS.

\subsection{The Coincidence Server}

There is currently a single SNEWS workstation running Linux at
Brookhaven National Laboratory, which serves as the coincidence
server.  Previously, we had had servers at LNGS and Kamioka, but moved
to BNL in fall of 2003, for ease of security and maintenance with the
resources available there.  The software has capabilities for dealing
with multiple servers, and more may be added in the future.  A second,
identical machine is kept running and in synch with the primary
server, so that immediate failover is possible in case of a problem with
the primary server.

The coincidence server remains behind the BNL firewall.  Only very
limited access to SNEWS subgroup members and the BNL system
administrator is permitted.  In addition, the server is housed in a
physically secure location.  If additional servers are added to the
network, they will be subject to similar security requirements.


\subsection{Coincidence Software}

The SNEWS software involves client and server programs which implement
a simple datagram exchange via socket, employing TCP/IP protocol, and
encrypted via OpenSSL.  The code is designed to be easily portable to
diverse operating systems.

The client software is provided to the individual experiments in the
form of a library of subroutines that may be called by an experiment's
supernova watch software to initiate a datagram transfer.  The package
also provides standalone tools for testing.

The server software runs in a standalone mode, and most of the time
simply waits to receive datagrams from the clients.  It maintains two
queues: a normal queue and a ``high-rate'' queue, for test alarms.
When an alarm datagram is received, it is placed on a queue according
to its flag (see Section~\ref{types}.)  One month's
worth of alarms are stored in the queue.  Received alarms are written
to disk, and are read in from disk if the server is stopped and restarted.

Every time an alarm is received, the last 24 hours' worth of alarms on the
queue is searched for a coincidence.  See Section~\ref{coinc} for detailed
coincidence conditions.

When a client initiates a connection, the server employs several
layers of checks to validate the origin of the datagram.  Only the IP
addresses of the client machines of the involved experiments are
allowed to submit packets.  In addition the client and server exchange
certificates which have been verified by the SNEWS Certificate
Authority, and the server rejects the connection if any check fails.

\subsection{SNEWS Shifts}

SNEWS subgroup members share shiftwork on a regular cycle.  Shift
duties include a check twice daily to ensure that the server is
running, that network connectivity is up, and that communication
capability is in good order.

Individual experiment alarm rates are monitored by SNEWS subgroup
members, so that any long term increase of rate 
over the 1/week limit may be addressed.

\subsection{SNEWS Operational Modes}

We have established a well-defined operational mode for SNEWS, which
we expect to develop in a series of managed transitions between
operational modes. For instance, new experiments or new coincidence
servers will be added or removed.  Each operational mode is identified
by a number and the date when it came into effect, and will specify in
detail the participants, the coincidence conditions, the alert
classifications, and the procedures for action in case of different
alarm conditions.  The following sections outline the conditions
for the operational mode we anticipate for the near future.

\subsection{SNEWS Packet Types and Flags}\label{types}

Each participating experiment may generate and send to the server
different types of alarm datagrams.
The alarm datagrams include a packet type, and a level flag.
The packet type can be PING, ALARM, or RETRACTION.
The level flag can be TEST, GOOD, POSSIBLE, RETRACTED or OVERRIDE.  
Datagrams having packet field values which do not 
belong to any of these categories are discarded by the server.

\noindent \underline{Packet Types}

\begin{itemize}
\item PING: Ping packets are used for test purposes only
and cause nothing more than a message printed to the coincidence
server log.

\item ALARM: Alarm packets contain information about individual
experiment alarms; what the server does with them depends on the
level flag.

\item RETRACTION:  Retraction packets contain information about
previously sent alarms to be retracted from the server's alarm queues.

\end{itemize}

\noindent \underline{Level Flags}

\begin{itemize}

\item TEST: This flag indicates a datagram packet intended for test use
as well as for any high-rate test mode.  

\item POSSIBLE : This flag indicates an alarm generated during
scheduled operations (i.e. maintenance, calibration, tests, etc.) or
other known anomalous conditions. It is up to each experimental
collaboration to set this flag inside the packet when appropriate.

\item GOOD: This flag indicates an alarm generated during
normal detector conditions.  

\item RETRACTED: This flag is set for retraction 
packets (note that this information is redundant---all packets of
RETRACTION type will be retracted regardless of level flag.)  

\item OVERRIDE: This flag indicates an alarm 
that has been confirmed as good.  
\end{itemize}

\subsection{Coincidence Definition}\label{coinc}

The general coincidence definition implemented in the coincidence code
may generate either of two types of alert: GOLD or SILVER.\\
A GOLD alert is generated if {\it all} of
following conditions (1 through 4) are met:

\begin{enumerate}

\item There is a 2 or more -fold coincidence (by UT time stamp) 
within 10 seconds, involving at least two different experiments. (The
time window refers to the maximum separation of any of the alarms in
the coincidence.)

\item At least two of the experiments involved
are at physically separated laboratories.  This condition is automatically
satisfied for the current operational mode.

\item Two or more of the alarms in the coincidence
are flagged as GOOD.  It is the responsibility of each participant
experiment to flag the alarm sent to the SNEWS server(s)
appropriately. The specific criteria for GOOD/POSSIBLE alarms
are locally defined by each experiment.
  
 \item For at least two of the experiments involved in the
 coincidence, the rate of good alarms for several past time intervals
 $\{T_i\}=\{$10 minutes, 1 hour, 10 hours, 1 day, 3 days, 1 week, 1
 month$\}$ preceding the first alarm of the coincidence candidate,
 must be consistent with the $\lambda_{\rm max}=$1/week
 requirement.\footnote{These intervals represent real time, not live
 time, since full live time information will not be available to the
 coincidence server.}  We define the precise condition as follows: if
 an experiment sent $\{n_i\}$ alarms in each of the last intervals
 $\{T_i\}$ before the first event of the coincidence, then the Poisson
 probabilities $\mathcal{P}_i$ for $n_i$ or more alarms in $T_i$,

$\mathcal{P}_i=\sum_{n=n_i}^{\infty}(\lambda_{\rm max}
T_i)^{n}e^{-\lambda_{\rm max} T_i}/n!$,

for each interval $T_i$, must each be greater than $\mathcal{P}_{thr}=0.5$\%.
This corresponds to the condition that each $\{n_i\}$ must be less
than $\{1,2,2,3,4,5,11\}$ for the preceding intervals $\{T_i\}$ for an
alarm to be GOLD.

\end{enumerate}

When the first criterion is satisfied, but at least one of the other
criteria is not satisfied, the generated alert is flagged as SILVER.
In this case the alert has to be checked by the individual
experiment collaborations before any public announcement. No alert
will be sent to the community by SNEWS until (and if) there is an
upgrade to GOLD.

\subsection{Rate-dependent GOLD Coincidence Suppression}\label{suppression}

\begin{figure}[t]
\hspace{2cm}
\begin{center}
\includegraphics[width=11 cm]{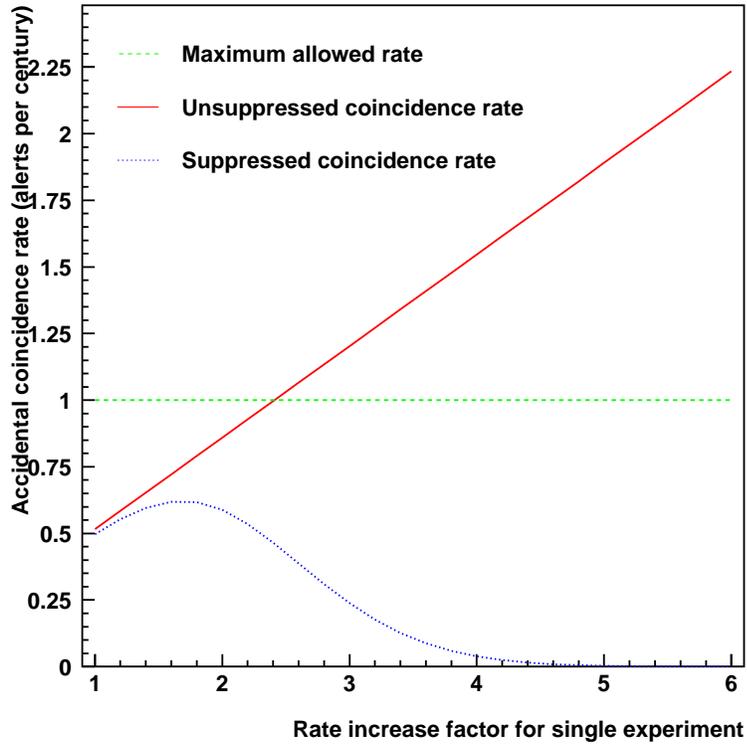}
\caption{\label{fig:as}
The green dashed line shows the maximum allowed
accidental coincidence rate (1 per century.)  The red line shows
the expected accidental coincidence rate of 2 out of 3 experiments (for
a 10~second coincidence window), under
the assumption that one of the three experiences an increase in rate by
a factor $f$ (and the others
maintain a 1 per week rate), as a function of rate increase factor $f$.  
The blue dotted line shows the expected coincidence rate after the past rate history
suppression has been applied.
 }
\end{center}
\end{figure}

The last criterion---demotion to SILVER based on past
rate history---deserves some additional discussion.  
The purpose of this criterion is to protect against short
term rate increases from one or more experiments.  The suppression
is effective: see Figure~\ref{fig:as}.  However it comes at
a slight cost: if one assumes a constant
Poisson background rate of 1~per week for all three experiments,
criterion~4 will result in demotion of about 4\% of true GOLD alerts
to SILVER, just due to Poisson fluctuations in the previous
time windows.  However, the protection
against unexpected increases in background rate is probably worth this
small loss (note that typically individual experiment alarm rates will
be less than 1 per week anyway.)  

\begin{figure}[htbp]
\begin{center}
\includegraphics[width=11cm]{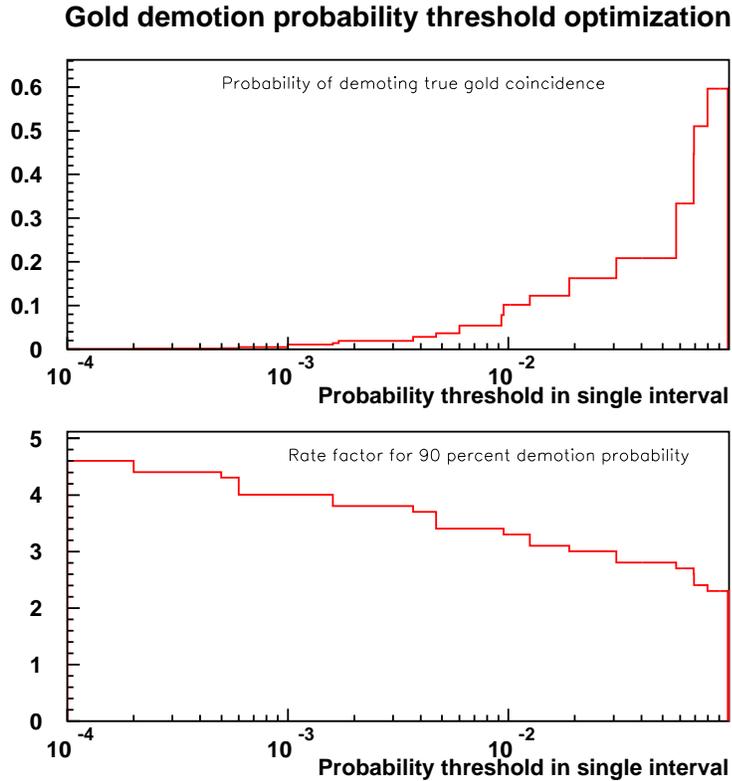}
\caption{\label{fig:criterion4}
Top plot: total probability of demoting a true GOLD coincidence as a function
of $\mathcal{P}_{thr}$, assuming three experiments with 1/week alarm rate.
Bottom plot: factor by which average rate of any
single experiment would
have to increase in order for 90\% of coincidences to be
demoted. The  $\mathcal{P}_{thr}$ chosen is 0.005.}
\end{center}
\end{figure}

One might
also worry that long term rate increases might cause increased
demotion of true GOLD to SILVER. 
We have evaluated the overall average rate increase 
from any single experiment that would result in
90\% of true coincidences being demoted.
Figure~\ref{fig:criterion4} shows the effect of changing
$\mathcal{P}_{thr}$.  The value of $\mathcal{P}_{thr}$ chosen was
0.5\%, which gives fairly low true GOLD suppression (4\%); and at this
threshold any overall single experiment rate increase of more than a
factor of 4 will result in demotion of 90\% of
coincidences.\footnote{ Note that long term average alarm rates will
be monitored by shiftworkers, and
\textit{subgroup
members will be notified if rates of their experiments exceed the
nominal 1 per week limit}, so any such rate increase will be
temporary.  Also note that if alarm rate increases have clearly been
corrected, by subgroup agreement on a case-by-case basis individual
alarms may be retracted after the fact, so as not to decrease true
GOLD coincidence efficiency.}

We feel this GOLD and SILVER scheme strikes the right balance
between danger of losing true coincidences due to too-stringent
criteria and danger of issuing false alerts to astronomers.

\subsection{Demotion and Promotion}

Although we hope to avoid ever being in the situation where retraction
of a GOLD alert is necessary, any experiment may reflag from GOOD (or
POSSIBLE) to RETRACTED its own alarm after data checking.  The server
will then automatically reevaluate and reissue the alert based on
alarms in the past day of its memory: the result may be still GOLD,
demotion to SILVER, or no alert at all.  For the latter case, the
SNEWS subgroup is notified, and a RETRACTED alert will be issued to the
same mailing list as for GOLD and posted on the public web page.

Experiments may also send OVERRIDE packets: a GOLD alert may also be
generated if condition 1 is satisfied and at least one alarm in the
coincidence is OVERRIDE and at least one is GOOD, regardless of
whether the other conditions are satisfied. This case allows an
override of past high-rate history demotion (or other conditions that
could tag an alarm as POSSIBLE) for a human-checked alert.

\begin{figure}[htbp]
\begin{center}
\includegraphics[width=8cm, bb= 40 120 539 820]{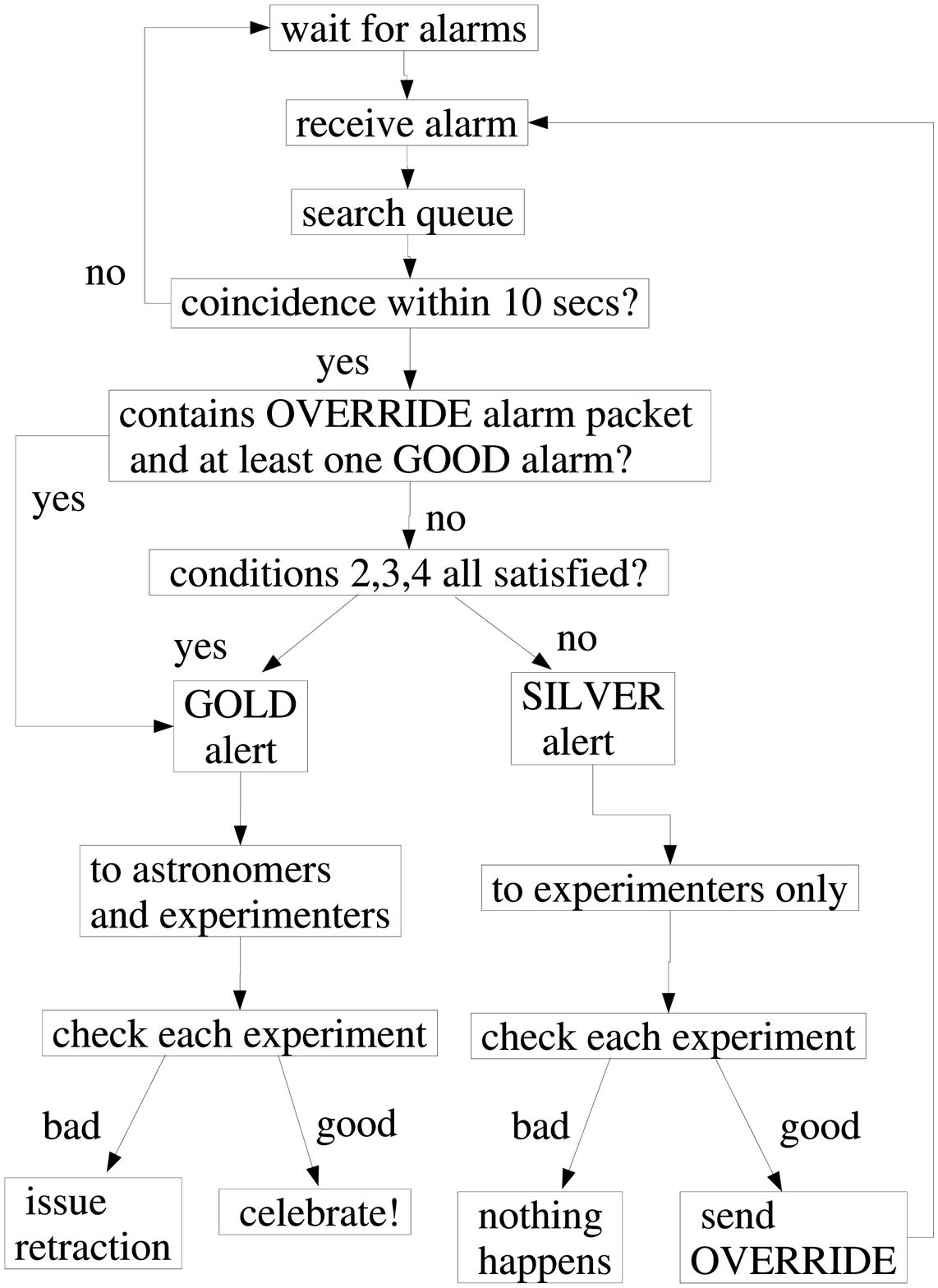}
\caption{\label{fig:flowchart} Flowchart summarizing the 
sequence of events and decisions that determine whether an
alert is GOLD or SILVER.}
\end{center}
\end{figure}

Figure~\ref{fig:flowchart} summarizes the sequence of events 
and GOLD vs. SILVER decisions.

\section{High Rate Test Results}\label{highrate}

During an approximately two-month period in April-June of 2001,
Super-K, LVD and SNO subgroup members performed a ``high rate test''
of the coincidence software.  The purpose was two-fold: first, to
check the robustness of the software and work out any remaining bugs;
and second, to increase confidence in our understanding of the
expected coincidence rates.
The test was highly successful.  

The idea of the high rate test was to lower the thresholds of the
experiments' supernova monitors such that coincidence
alerts increased to a non-negligible rate, due to the Poissonian nature
of the data.  Each experiment set its
supernova monitor burst search parameters to yield a
test alarm rate somewhere in the range of 10-100/day.  In addition, in
order to
increase artificially the coincidence rate, the coincidence window was
increased from its standard value of 10~seconds to 400~seconds.

The individual experiment alarm and coincidence rates were somewhat
non-stationary, which was not unexpected.  The results were analyzed
via a ``time-shift'' method (see below) to show that alarms were uncorrelated,
and that recorded coincidences were consistent with expected rates.

\begin{figure}[htbp]
\hspace{2cm}
\begin{center}
\includegraphics[width=11cm]{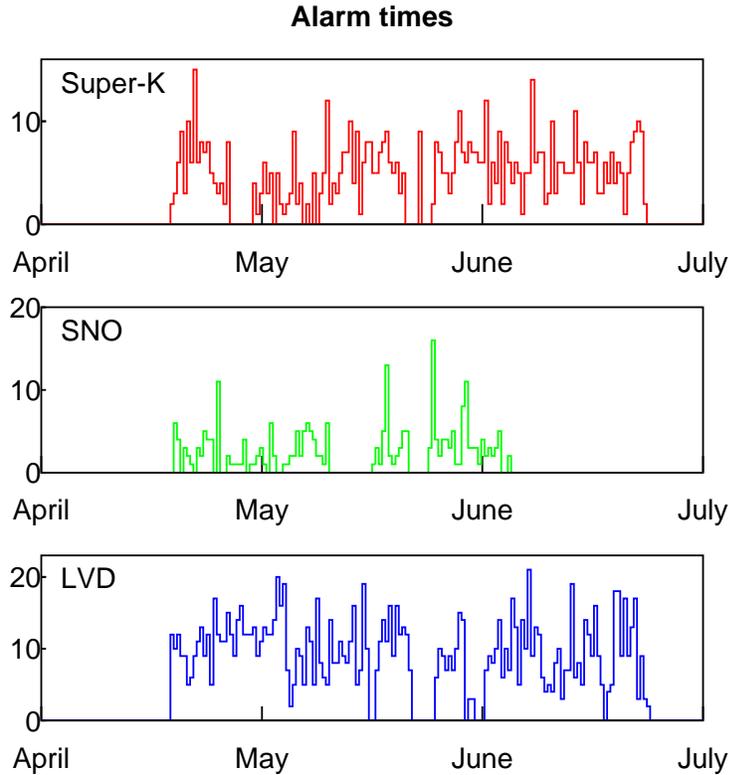}
\caption{\label{fig:alarms1}
Number of alarms received from the individual experiments, 
plotted in 11 hour bins.}
\end{center}
\end{figure}

\begin{figure}[htbp]
\hspace{2cm}
\begin{center}
\includegraphics[width=11cm]{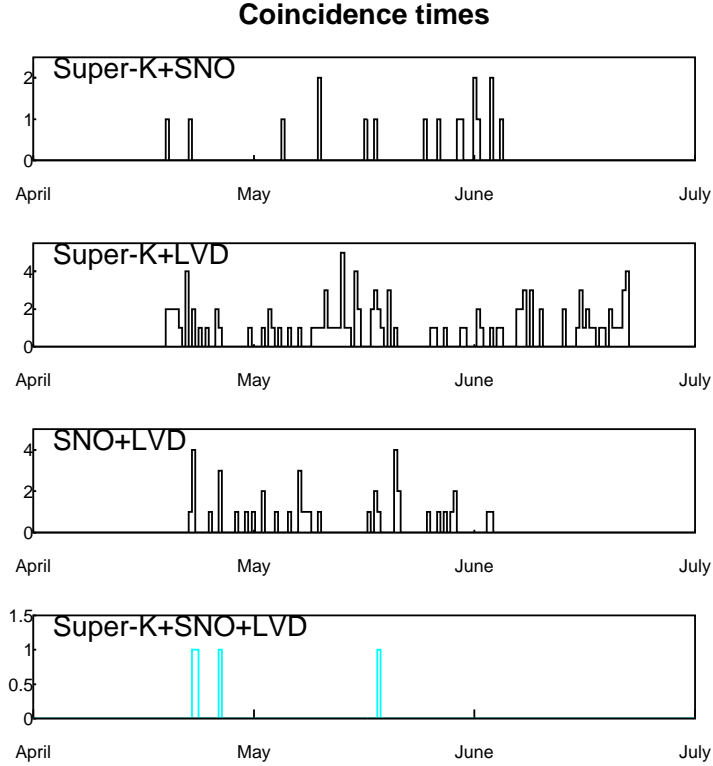}
\caption{\label{fig:coinc}
2- and 3-fold coincidences, plotted in
11 hour bins. ``Unique'' coincidences only are shown.}
\end{center}
\end{figure}

The alarms received as a function of time, and coincidences as a
function of time, are shown in Figures~\ref{fig:alarms1}
and~\ref{fig:coinc}.  The numbers of individual alarms, and numbers of
2- and 3-fold coincidences are shown in Table~\ref{tab:compare}.  The
rates are roughly constant over most periods, although there is
clearly some ``burstiness''.  

Experiments can be ``dead'' to SNEWS for many reasons:  actual detector
deadtime, online supernova monitor problems, network problems,
or coincidence server problems.
To estimate the deadtime, we used the data themselves and counted
improbably long gaps as deadtime.   This method automatically takes
into account dead time from all causes.

After removing the dead intervals, we calculated the overlapping
live periods for each pair of detectors, as well as the 3-experiment
overlap time.  Note that there are some dead intervals common to all
detectors due to network trouble; in particular the period from May
22-25 represents a problem with network availability to the test
server at the Kamioka site (note that we expect very high uptime at
the current BNL server site.)

Based on these known alarm rates and live periods, we then calculate
expected accidental coincidence rates.  For the purpose of
comparison with expected coincidence rates, we have calculated the
``raw'' number of coincidences from the individual experiment alarms
arriving at the server: the number of raw coincidences is defined as
the number of times the individual experiment alarms are separated by
a maximum test time window of $\tau$=400~s.  

Note that in the real
case, according to the rate history-based demotion algorithm described
above, the coincidence server will suppress redundant alerts.  If
there is more than one alarm from a given experiment within the time
window, the server will send only one GOLD alert, corresponding to the
coincidence of the first alarm from each experiment (assuming all
conditions are satisfied) -- whereas multiple ``raw'' coincidences
would be counted for this cluster.  Subsequent coincidences would be
demoted to SILVER.  The number of ``unique'' coincidences is the
number of coincidences with different first alarm times.

The final column of Table~\ref{tab:compare}
shows both ``raw'' and ``unique'' (in parentheses) numbers of
coincidences.  Figure~\ref{fig:coinc} shows ``unique'' coincidences.
The coincidence server output was checked 
to verify consistency with the calculated raw coincidences.

For stationary, uncorrelated Poisson point processes,
the rate of $N$-fold coincidences between $N$ detectors
is given by

\begin{equation}\label{eq:Nfold}
\lambda_{\rm coinc}= N\left( \frac{\tau^{N-1}}{T_{\rm obs}^N}\right)
\prod_{i=1}^N \mathcal{N}_{i},
\end{equation}

where $\tau$ is the coincidence window (the maximum separation of
events for a coincidence), $T_{\rm obs}$ is the total common
observing time, and $\mathcal{N}_i$ is the number of events observed
by the $i$th detector.  For example, for a 2-fold coincidence between
detectors $i$ and $j$, the expected number of coincidences is
2$\mathcal{N}_i \mathcal{N}_j \tau /T_{{\rm obs}~ij}$.





The uncertainties on the expectated rate values are calculated by
propagating the uncertainties on the live time.

However, equation~\ref{eq:Nfold} is strictly valid only for stationary
processes, and this assumption is clearly violated in our case (see
Figure~\ref{fig:alarms1}.)  Therefore we take a different approach to
calculate expected coincidence rates: to predict more generally the
number of accidental coincidences from these non-stationary alarm
sequences, we apply a ``time-shift'' method~\cite{Amaldi, Prodi,
Allen}: for any pair of detectors, we shift all of one experiment's
alarm time values by an offset $\Delta t$, and determine the number of
coincidences $n_c$ for that time offset value.  This procedure is
repeated for many values of $\Delta t$; the mean and standard
deviation of the distribution of $n_c$ values then gives both the
expected number of observed coincidences and its expected spread,
which we then compare with the observed number of raw coincidences.
Similarly, we time-shift one of the three experiment's alarm time
series by $\Delta t$ to determine the expected 3-fold coincidence
rate.

\begin{figure}[htbp]
\begin{center}
\includegraphics[width=11cm]{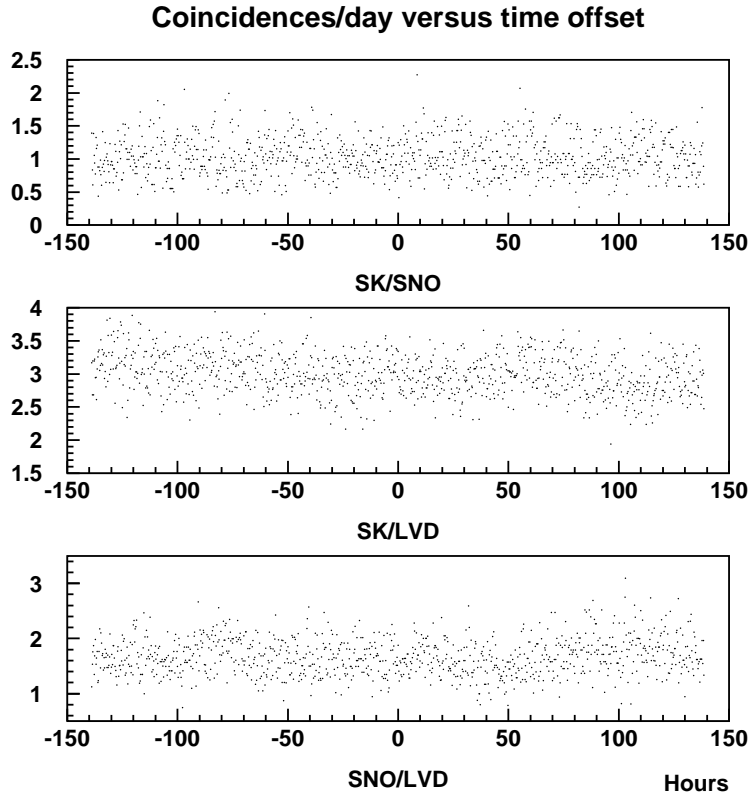}
\caption{\label{fig:ratevsoffset}
Rate of 2-fold coincidences for each experiment pair,
as a function of time offset in hours.  The rate was determined
using overlap live time after the time offset.}
\end{center}
\end{figure}

The plot of $n_c$ versus time offset value should be flat, and show no
spike at zero (or any other) offset, if there are no correlations
between the different experiments' alarm times.  The results of this
analysis for 2-fold coincidences are shown in
Figure~\ref{fig:ratevsoffset} (a similar plot, although with lower
statistics, results for 3-fold coincidences.)  We use time shifts
ranging from -150 to 150 hours at 1000 second intervals.  Live time is
taken into account in the time-shifted sample by shifting the offset
experiment's live period by the same offset and then re-evaluating the
overlap time. The mean and RMS values of the resulting shifted
coincidence rates are used to determine the expected number of
coincidences for each combination in Table~\ref{tab:compare}.

Table~\ref{tab:compare} shows the expected and observed numbers of
events.  The expected numbers of coincidences from
equation~\ref{eq:Nfold} do not exactly match the expected numbers from the
time-shift method, even considering live time estimate uncertainty.
Presumably this is due to the somewhat non-stationary nature of the
alarm sequence.  The number of observed coincidences do match the
time-shift expectations well within the expected spread.  In addition,
the time shift plots show no evidence of correlations between
experiments, as expected.

\begin{table}[h]
\centering
\begin{tabular}{||c|c|c|c|c|c|| }
\hline\hline
Experiment & Common &  SK/SNO/LVD & $N_{\rm coinc}$ & $N_{\rm coinc}$ & $N_{\rm coinc}$ \\
Combination  &  live time & alarms &expected  &expected & observed  \\ 
                       &  (days) &               & (eqn) & (shift) & raw (unique)  \\ \hline\hline
SK/SNO    &  24.1$^{+1.1}_{-0.5}$  & 334/187/-  & 24.1$^{+1.0}_{-2.2}$   & 24.9 $\pm$ 7.0& 30 (17)\\ \hline
SK/LVD    &  44.6$^{+1.1}_{-0.9}$  & 576/-/1025  & 122.6$^{+4.9}_{-5.8}$  & 133.8 $\pm$ 13.7 & 149 (112)\\ \hline
SNO/LVD   &  27.7$^{+0.7}_{-0.6}$  &  -/189/646   & 40.8$^{+1.6}_{-2.0}$   & 46.4 $\pm$ 9.2 & 52 (41) \\  \hline
SK/SNO/LVD & 19.6$^{+1.1}_{-0.6}$  & 276/144/431  & 2.9$^{+0.3}_{-0.5}$ & 4.2 $\pm$ 2.9 & 4 (4) \\ \hline\hline

\end{tabular}
\caption{\label{tab:compare}Alarm and coincidence
summary.  The first column indicates the 2 or 3-fold experiment
combination.  The second column gives the overlap live time for that
combination with estimated uncertainties.  The next column gives the
numbers of alarms from the experiments which are within the overlap
live time.  The fourth column gives the number of expected
coincidences according to equation~\ref{eq:Nfold}; the error reported
is the systematic error only, from uncertainty in the live time
estimate.  The fifth column gives the expectation and its RMS based on
the time shift method.  The final column gives the observed number of
coincidences during the test; the number outside parentheses indicates
the ``raw'' number of coincidences defined as the number of alarms
with maximum time separation of 400~seconds (to be compared
to the predictions); the number in parentheses
is the number of ``unique'' coincidences tagged by the coincidence
server (see text.)  }

\end{table}

Although these somewhat non-stationary data, taken at lowered
threshold, do not necessarily imply that rates will also be
non-stationary when thresholds are raised and running conditions are
normal, one can never be completely sure that individual experiment
rates will not increase unexpectedly.  This is the motivation for the
rate-dependent GOLD suppression scheme of section~\ref{suppression}.

The coincidence server now has capability for continuous
high rate testing, using tagged TEST alarms in parallel with normal
alarms.

\section{The Alert to the Astronomical Community}\label{alert}

At the supernova early alert workshop of 1998, the conclusion from the
astronomer working group~\cite{workshop} was that ``the message will
spread itself'' and that SNEWS will need to do no more than send out
emails to as many astronomers as possible.  SNEWS maintains a  mailing
list of interested parties, including both professionals and amateurs,
to be alerted in the case of a coincidence.   

In an ideal case, the coincidence network provides the astronomical
community with an event time and an error box on the sky at which
interested observers could point their instruments.   In a realistic 
case, the size of the error box is dependent on the location of the 
supernova and the experiments which are online, and may be very large
(and at this time will not be available in the initial alert message.)
However, members of the mailing list with wide-angle viewing capability
(satellites, small telescopes) should be able to pinpoint 
an optical event quickly.   Although an unknown fraction of galactic
supernovae will be obscured by dust, many will be visible to amateurs
with modest equipment.

Regardless of the quality of neutrino pointing available, however, the 
advance warning alone gives observers of all kinds valuable time to get 
to their observatories and prepare to gather data as soon as an accurate 
position is determined. 

A Target of Opportunity proposal for the Hubble Space Telescope, 
``Observing the Next Nearby Supernova'', aiming to take advantage of early 
supernova light based on an early warning, was approved~\cite{HST} for 
Cycle 13 and was operational for Cycles 8 through 12.   

\subsection{Amateur Astronomers}

The large pool of skilled and well-equipped amateur astronomers is
also prepared to help locate a nearby supernova.  The editors of {\it
Sky \& Telescope} magazine have set up a clearinghouse for amateur
observers in search for first light (and a precise optical position as
early as possible)~\cite{skyandtel}, via their AstroAlert
service~\cite{astroalert}.  This was started by former
editor-in-chief Leif Robinson, and has the continued support of
current editor-in-chief Rick Fienberg.   In collaboration with the American 
Association of Variable Star Observers, they have developed a set of 
criteria for evaluating amateur responses to an alert, so that a 
reliable precise position can be disseminated as early as possible.  For 
instance: there must be at least two consistent reports, demonstrated 
lack of motion, lack of identification with known asteroid and variable 
star databases, variability consistent with supernova light curves and, 
if the information is available, a spectrum consistent with known 
supernova types. 
 
On February 14 2003, {\it Sky \& Telescope} performed a test for
amateurs.  A transient target (the asteroid Vesta at a near-stationary
point in its retrograde loop) was selected, which at the time was
about magnitude 6.7.  {\it Sky \& Telescope} issued an alert (very
carefully tagged as a test) to their mailing list, with a given
13-degree uncertainty radius. They received 83 responses via the web
response form, and more by email.  The responses were of world-wide
distribution, and although many observers experienced poor conditions,
six were successful in identifying the target.  From this experience,
they have suggested refinements to optimize amateur astronomer
strategy.  A second test is planned soon, and should be a regular
occurrence.

\subsection{SNEWS Alerts}

We maintain two alert mailing lists which will be sent to
automatically by the SNEWS coincidence software in the case of an
alert.  The first is the GOLD alert list, which includes all
astronomers who have signed up, including {\it Sky \& Telescope} and the HST
astronomers, and is to be an \textit{automated} alert.  The second
mailing list will be for SILVER alerts, and is to be sent to neutrino
experimenters only.  These alerts will be checked out by shiftworkers
at their respective experiments before an alert is issued; each
experiment is responsible for making sure the SILVER alert messages
reach shiftworkers.  Each experimental collaboration defines its own
protocol for acting on a SNEWS SILVER or GOLD alert.

For both SILVER and GOLD cases, a message containing the following
information:

\begin{itemize}
\item UTC time of the coincidence,
\item all detectors involved in the coincidence, and
\item the types of alarms (GOOD, POSSIBLE) for each experiment involved
in the coincidence
\end{itemize}

will be automatically sent by the server to the SNEWS subgroup
members.   The information may also be posted to a restricted SNEWS
subgroup page for SILVER, and a public page for GOLD.

To allow the confirmation of a SNEWS alert as really coming from SNEWS,
any alerts will be public key signed using the SNEWS key.  This key has
the ID\#~68DF93F7, and is available on the network of public PGP
keyservers such as \texttt{http://pgp.mit.edu/}

Note that there is no restriction on individual experiments making any
announcement based on individual observation in the case of absence of
a SNEWS alert, SILVER or GOLD, or preceding or following any SNEWS
alert message.  Any individual experiment may publicly announce a
supposed supernova signal following a dispatched SILVER alert which
has not yet been upgraded to GOLD.  In this case the information that
a previous SILVER alert from the SNEWS server(s) has been received
should be cited.

\section{Status and Future Prospects}\label{future}

At the time of this writing, SILVER alerts only between Super-K and
LVD are activated.  We are working towards having the operational mode
described in this paper to be activated in the very short term,
comprising automated GOOD alarms from Super-K and LVD, but automated
POSSIBLE alarms only from SNO, such that SNO will participate in a GOLD alert
only if at least two other experiments' GOOD alarms are present.

We also expect SNEWS to incorporate more galactic-supernova-sensitive
neutrino detectors over the next few years.  In addition, we may
expand the network of servers with additional secure sites.

\section{Summary}

In summary, several supernova neutrino detectors are now online.  
If a stellar core collapse occurs in our Galaxy, these 
detectors will record signals from which a wealth of physical and 
astrophysical information can be mined. 
 
An early alert of a gravitational collapse occurrence is essential to 
give astronomers the best chance possible of observing the physically 
interesting and previously poorly observed early turn-on of the 
supernova light curve.  A coincidence of several neutrino experiments 
is a very powerful technique for reducing ``non-Poissonian'' 
false alarms to the astronomical community, in order to allow a
prompt alarm.  We have implemented such a system, currently 
incorporating several running detectors: LVD, SNO and Super-K.
We expect to expand the network in the near future, and move
to a more automated mode in the near future.

\ack{ We would like to thank the members of the Super-K, SNO and LVD
experimental collaborations, without whose hard work the SNEWS project
would not be possible.  We thank Sam Aronson and Tom Schlagel of
Brookhaven National Laboratory for providing a secure home for the
server.  We are grateful to John Bahcall, Barry Barish,
John Beacom, Janet Conrad, Ed Kearns,
Janet Mattei, Jim Stone, and Larry Sulak for advice and support; others
from the early days of SNEWS are Adam Burrows,
Robert Kirschner, and Mark Vagins.  We
also thank the members of the SNEWS working group from other
experiments.  SNEWS is funded by National Science Foundation awards
PHY-0303196 and PHY-0302166.  }


\section*{References}
\begin{thebibliography}{99}

\bibitem{Hirata1} Hirata K S {\it et al.} 1987 {\it Phys. Rev. Lett.} {\bf 58} 1490 

\bibitem{Bionta} Bionta R M {\it et al.} 1987 {\it Phys. Rev. Lett.} {\bf 58}  1494 


\bibitem{Bratton}
Bratton C B {\it et al.} 1988 {\it Phys. Rev.} D {\bf 37}  3361


\bibitem{Hirata2}
Hirata K S {\it et al.} 1988 {\it Phys. Rev.} D {\bf 38}  448 

\bibitem{Alekseev} Alekseev E N {\it et al.} 1987 {\it JETP. Lett.} {\bf 45}
589 

\bibitem{Aglietta}Aglietta M {\it et al.} 1987 {\it Europhys. Lett.} {\bf 3}
1315 

\bibitem{Burrows} Burrows A {\it et al.} 1992 {\it Phys. Rev.} D {\bf 45}
3361

\bibitem{Thompson} Thompson T {\it et al.} 2003 {\it Astrophys. J.} {\bf 592} 434

\bibitem{Mezzacappa} Liebendorfer M {\it et al.} {\it Preprint}  astro-ph/0207036

\bibitem{snwatch} Cline D B 1990  \textit{Proceedings of the Supernova Watch Workshop (Santa Monica)}

\bibitem{pointing} Beacom J and Vogel P 1999 {\it Phys. Rev.} D {\bf 60} 033007
\item[] (Beacom J and Vogel P 1999 \textit{Preprint} astro-ph/9811350)

\bibitem{snfreq} Tammann G A {\it et al.} 1994
{\it Ap. J. Supp.} {\bf 92} 487 

\bibitem{macro} Ahlen S {\it et al.} 1992 {\it Astropart. Phys.} {\bf 1} 11
\item[] Ambrosio M {\it et al.} 1998 {\it Astropart. Phys.} {\bf 8} 123-133

\bibitem{nu2k} Scholberg K 2000 {\it Nucl. Phys. Proc. Suppl.} {\bf 91} 331
\item[] (Scholberg K 2000 \textit{Preprint} hep-ex/0008044)

\bibitem{LVDref} Aglietta M {\it et al.} 1992 \textit{Nuov. Cim.} A {\bf 105}, 1793

\bibitem{superk} Fukuda S {\it et al.} 2003 \textit{Nucl. Instrum. Meth.} A {\bf 501} 418

\bibitem{SNO} Boger J {\it et al.} 2000 {\em Nucl. Instrum. Meth.} A {\bf 449} 172
\item[] (Boger J {\it et al.} 2000 \textit{Preprint} nucl-ex/9910016)

\bibitem{Amanda} Jacobsen J E, Halzen F and Zas E 1994  \textit{Phys. Rev.} D {\bf 49} 1758
\item[] Jacobsen J E, Halzen F and Zas E 1996 \textit{Phys. Rev.} D {\bf 53} 7359
\bibitem{icarus} Bueno A, Gil-Botella I and Rubbia A 2003 \textit{Preprint} hep-ph/0307222 

\bibitem{miniboone} Sharp M {\it et al.} 2002 \textit{Phys. Rev.} {\em D66} 013012 
\item[] (Sharp M {\it et al.} 2002 \textit{Preprint} hep-ph/0205035)


\bibitem{borexino} Cadonati L {\it et al.} 2002 \textit{Astropart. Phys.} {\bf 16} 361 
\item[] (Cadonati L {\it et al.} 2002 \textit{Preprint} hep-ph/0012082)

\bibitem{omnis} R. N. Boyd {\it et al.} 2003 \textit{Nucl. Phys.} A {\bf 718} 222

\bibitem{uno} C. K. Jung 2000 \textit{Preprint} hep-ex/0005046 

\bibitem{whitepaper} Homestake Collaboration 2003 \textit{NUSEL Science Book} 
\textit{Preprint} nucl-ex/0308018 

\bibitem{ness} Scholberg K 2003
\texttt{http://mocha.phys.washington.edu/$\sim$int\_talk/WorkShops/Neutrino02/\\Working\_Groups/People/Scholberg\_K/scholberg\_thurs\_solarstellarwg.pdf}

\bibitem{gravwave} Hughes S A \textit{et al.} 2001 \textit{Preprint} astro-ph/0110349  
\bibitem{igec} Astone P \textit{et al.} 2003  \textit{Phys. Rev.} D {\bf 68} 022001
\item[] (Astone P \textit{et al.} 2003 \textit{Preprint} astro-ph/0302482)

\bibitem{Arnaud} Arnaud N {\it et al.} 2000 \textit{Phys. Rev.} D {\bf 65} 033010\item[] (Arnaud N {\it al.} 2000 \textit{Preprint} hep-ph/0109027)

\bibitem{sk} Fukuda Y 2003 {\it Proc. of Origin of Matter and Evolution of Galaxies 2000 (World Scientific)} 209
\item[] Super-K collaboration, in preparation

\bibitem{sno} Virtue C 2001 {\it Nucl. Phys. Proc. Suppl.} {\bf 100} 326 
\item[] (Virtue C 2001 \textit{Preprint} astro-ph/0103324)
\item[] SNO collaboration, in preparation

\bibitem{lvd} Fulgione W 1999 {\it Nucl. Phys. Proc. Suppl.} {\bf 70} 469 

\bibitem{sk_solar} Fukuda S {\it et al.} 2001 {\it Phys. Rev. Lett.} {\bf 86} 
5651
\item[] (Fukuda S {\it et al.} 2001 \textit{Preprint} hep-ex/0103032)

\bibitem{Amaldi} Amaldi E {\it et al.} 1989  \textit{Astron. Astrophys.} {\bf 216} 325

\bibitem{Prodi} Prodi G A {\it et al.} 2000 \textit{Int. J. Mod. Phys.} D {\bf 9} 237
\item[] (Prodi G A {\it et al.} 2000 \textit{Preprint} astro-ph/0003106)

\bibitem{Allen}Allen Z A {\it et al.} 2000  \textit{Phys. Rev. Lett.} {\bf 85}
5046 
\item[] (Allen Z A {\it et al.} 2000 \textit{Preprint} astro-ph/0007308)


\bibitem{workshop}\texttt{http://hep.bu.edu/$\sim$schol/workshop/workshop\_summary.html}

\bibitem{HST} Bahcall J (P.I.) 2003 \textit{Observing~the~next~nearby~supernova} HST proposal 10264 \texttt{http://presto.stsci.edu/public/propinfo.html}

\bibitem{skyandtel}Robinson L J August 1999 {\it Sky~\&~Telescope} 30

\bibitem{astroalert}\texttt{http://SkyandTelescope.com/observing/proamcollab/astroalert/}


\end{thebibliography}
\end{document}